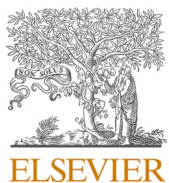

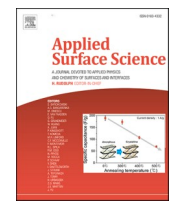

Full Length Article

# Thermal stability of interfacial mixed layers in c-Ni/a-Zr multilayer during annealing: Structural and magnetic properties

Debarati Bhattacharya [a,d,*], Vijay Karki [b,d], Surendra Singh [a,d], T.V. Chandrasekhar Rao [c,d]

[a] Solid State Physics Division, Bhabha Atomic Research Centre, Mumbai 400085, India
[b] Fuel Chemistry Division, Bhabha Atomic Research Centre, Mumbai 400085, India
[c] Technical Physics Division, Bhabha Atomic Research Centre, Mumbai 400085, India
[d] Homi Bhabha National Institute, Anushakti Nagar, Mumbai 400094, India

ABSTRACT

Annealing of crystalline multilayers composed of two miscible elements usually causes interfacial mixing of the constituent atoms, possibly leading to the formation of binary alloys at the interfaces. Magnetron sputtered c-Ni/a-Zr multilayers deposited at room temperature were vacuum annealed isochronally at 200 °C and 400 °C to observe thermal stability of the interfaces constituting the crystalline/amorphous multilayers. Control over the interfacial behaviour can aid the formation of materials with novel properties. The resultant changes in structural and magnetic properties of the multilayers were investigated in detail through X-ray Reflectivity, Polarized Neutron Reflectivity, SIMS and SQUID-based magnetization measurement techniques. The compositional variations in the amorphous mixed layers formed at the Ni-on-Zr and Zr-on-Ni interfaces due to atomic transport, were carefully observed as a function of annealing. Interface widths proceeded to increase with annealing at the expense of the Ni and Zr layers. The Ni-on-Zr interface was seen to be unstable at both temperatures; but the overall response of the interfaces to atomic diffusion was more pronounced when the multilayers were annealed at 400 °C. Under conditions of supersaturation of atoms within restricted spaces of the interfacial layers and the limited availability of components within the multilayer, unexpected demixing effects were observed at both the interfaces. A large increase in magnetic moment obtained after annealing at 400 °C, was attributed to the high densification of the Ni layers as well as to the incorporation of Ni crystallites into the Zr-on-Ni interface layers.

## 1. Introduction

It is well established that thermal annealing of Ni/Zr crystalline multilayer structures can induce alloying and interlayer growth across the interfaces [1–7]. This propensity is due to the highly negative enthalpy associated with mixing of Ni and Zr [8]. The reaction temperatures in the Ni-Zr system are much lower than melting points of either of the reactants. This aspect endows a thermal study of Ni/Zr layered structures with wide possibilities, thereby making it very interesting. Apart from structural changes these multilayers also display interesting variation in magnetic properties with annealing [9]. The interfacial layers produced during thermal annealing of Ni/Zr crystalline multilayers are known to be predominantly amorphous [10], since these regions are kinetically constrained against the nucleation of crystalline structures. It is the wide inequality in diffusivities of Ni and Zr atoms that presents a barrier to the growth of crystalline compounds in the intermediate layers both in as-deposited and annealed states. However, the formation of crystalline intermetallic structures on annealing Ni/Zr crystalline multilayers have been reported at high temperature annealing through self-propagating reactions [3] and also beyond a certain width of the initial amorphous layers [4]. Extensive literature reports on solid state amorphization reactions in these multilayers conclude that Ni is the dominant moving species in this system [6,11,12]. The asymmetrical diffusion features of this system have been documented through MD simulation studies as well [13,14]. As the growth of the interfacial amorphous layer proceeds, further passage of the Ni atoms is slowed down [6,15,16]. The movement of Ni in amorphous alloy layers is then restricted through short-circuit diffusion paths [17].

Amorphous alloys of the Ni-Zr system in thin film form are a part of an important class of materials termed as thin film metallic glasses (TFMGs). The current importance of Ni-Zr TFMGs is evident from the reports of their diverse use; as in hydrogen purification [18], as potential biomaterials [19] and in nuclear reactor engineering applications [20]. The properties of these alloys can be tuned through detailed investigations of the nature of diffusion leading to their formation. The study of Ni/Zr multilayer interfaces offers a viable opportunity to determine the performance of such composite materials. Diffusion kinetics in amorphous alloys formed from precursor multilayers have been reported in a recent study of Cu-Zr and Ni-Zr alloys, which belong to the same family of early–late transition metal alloys. [21]. The Ni-Zr system is characterized by both strong miscibility and asymmetric diffusion of its components. Both these factors contribute to the rapid intermixing between Ni and Zr in multilayers and also to the build up of large stresses in this system due to unbalanced diffusional flow [22,23]. It has also been noted that the mechanical strength of these materials is related to this interdiffusional strain. The relation between interdiffusion and mechanical properties has been well demonstrated through simulation studies on Ni-Zr alloy systems [24,25] and recently through experiments as well [26]. The mechanical properties of Ni/Zr multilayers and Ni-Zr TFMGs dictates their applicability in various technological fields. Since TFMGs are amorphous they lack grain boundaries and therefore plastic deformation in these materials is restricted to shear bands. However, the utility of TFMGs in high strength structural applications is limited by shear band instability. This can be overcome by alternating amorphous and crystalline layers as has been demonstrated for several technologically important materials. [27]. Further, multilayers formed with alternating crystalline and amorphous layers of various materials also display superior properties, which can be exploited for potential engineering applications [28,29]. A huge amount of literature reports to date exists regarding the MD simulations for various mechanical properties of amorphous/crystalline nanofilms of the Cu-Zr system [30]. Such studies pertaining to the Ni-Zr system are conspicuous by their absence. The reported studies on Ni/Zr multilayers were all performed with crystalline components. Given the fact that Ni-Zr TFMGs have versatile applications, a better insight into the tuning of their properties through detailed studies about the nature of diffusion at the interfaces of a Ni/Zr multilayer are required.

The subject of the present work is the study of compositional variation of interfacial structure and associated magnetic properties in thermally annealed crystalline-Ni/amorphous-Zr (c-Ni/a-Zr) multilayer samples. The variations of interfacial atomic concentrations at the Ni-on-Zr and Zr-on-Ni interfaces due to thermal annealing were investigated in detail. Annealing temperatures used were well below the crystallization temperature of Ni-Zr alloys to allow a detailed investigation of diffusion in amorphous interfacial layers. The effect of the concomitant thermal stability of the interfacial mixed layers on the structural and magnetic properties of the multilayer were examined through X-ray Reflectivity (XRR), Polarized Neutron Reflectivity (PNR), Secondary Ion Mass Spectrometry (SIMS) and SQUID measurement techniques.

## 2. Materials and methods

Ni/Zr multilayers were deposited on Si substrates by D.C. magnetron sputtering in the form of 5 bilayers of nominal thickness Ni (145 Å)/Zr (155 Å) with a periodicity of 300 Å. The deposited Ni layers were crystalline and the Zr layers were amorphous. Details of the home-built 3-gun DC/RF magnetron sputter deposition unit for growing these multilayers can be found elsewhere [31]. The ultimate vacuum in the chamber was $3.8 \times 10^{-7}$ mbar and during deposition the vacuum level was maintained at $7 \times 10^{-3}$ mbar using a 20 sccm dynamic flow rate of Ar gas. Si substrates were loaded in the vacuum chamber through a load lock system to minimize contamination and held at 75 mm above high purity (~99.99%) Ni and Zr metal sputter targets fixed in a confocal geometry. Substrates were rotated during deposition to ensure lateral uniformity in thickness. The targets were presputtered before the deposition of multilayers to remove possible surface contaminants. In order to deposit sequential layers, the values of D.C. power coupled to each sputter gun were set at 16 W for Zr target and 25 W for Ni target. The target not in use for the particular layer being deposited was kept protected with a mechanical shutter operated from outside the vacuum chamber. Sputter deposition rates used for each target were: 0.1 Å/sec (Zr) and 0.6 Å/sec (Ni). Film thicknesses could thus be controlled through time of deposition. Room temperature (R.T.) as-deposited multilayer samples were reinserted into the same vacuum chamber for annealing treatments using an in-built substrate heater. The vacuum level during annealing was maintained at $10^{-5}$ mbar, which ensured that the films were free from adsorbed oxygen. The as-deposited multilayers are also referred to as R.T. in this paper. The multilayers were annealed in vacuum at 200 °C and 400 °C for one hour each. The thickness, density and roughness corresponding to the Ni and Zr layers, as well as that of the Ni + Zr mixed interface layers formed between these two layers were determined through analyses of XRR and PNR data. The interface layer formed when Ni is deposited on Zr is referred to as "Ni-on-Zr" in the text and the interface layer between Zr and Ni layers is termed as "Zr-on-Ni".

Specular XRR and grazing incidence XRD (at 1° incidence) measurements on the as-deposited and annealed films were performed in parallel beam geometry using a powder diffractometer with a Cu $K_\alpha$ ($\lambda$ = 1.54 Å) radiation source. The sample position was fixed and scanning was achieved by synchronized movement of the X-ray source and detector units. Care was taken so that the same critical alignment was used for both measurement modes to enable standard normalization procedures for each deposited sample. PNR measurements on each multilayer sample were performed in 2 kOe external magnetic field and recorded in specular mode at Dhruva reactor, India [32]. The data were analyzed using the same structural models used for XRR data analyses, in order to determine stoichiometry (composition) and magnetic structure. The SIMS technique was used to determine the depth distribution of elemental Ni and Zr, as well as molecular Ni-Zr species present at each interface layer of the as-deposited and annealed multilayer films with a Cameca IMS-7f secondary ion mass spectrometer. This technique was earlier found to be very useful to gauge mixed layer compositions [33]. A Cs$^+$ primary ion beam with beam current of $21 \pm 0.5$nA and impact energy of 5 keV was rastered over a scan area of 250 μm × 250 μm. A field aperture of 400 μm diameter was selected at the centre of the raster area in order to remove crater edge effects, which provided an analysis region of diameter ~33 μm. The analyses were carried out using MCs$^+$ secondary ion detection mode (M being the element of interest), in order to effectively reduce the matrix effects especially in the mixed layers [34]. Pressure in the analysis chamber was maintained at ~9 × 10$^{-9}$ mbar and mass resolution (m/dm) of ~400 was used to reduce isobaric mass interferences, arising from the sample or from background gases present in the analysis chamber. Changes in magnetic nature of the as-deposited and annealed multilayers were investigated through measurements with a Quantum Design SQUID magnetometer (model MPMS). The sample surfaces were aligned parallel to the direction of applied magnetic field in both field dependent and temperature dependent magnetization measurement modes. Field dependent magnetization for all the samples were recorded at both 5 K and 300 K.

## 3. Results

XRR measurements were made at grazing angles of incidence θ on the sample surface, so that the specularly reflected X-ray intensities could be recorded perpendicular to the surface as a function of momentum transfer q = (4π/λ)sin θ in Å$^{-1}$. The high depth sensitivity of this technique is a unique advantage used to probe differences in refractive indices of adjacent layers in stratified structures. This enables the collection of structural information as a function of depth along the surface normal to the sample. The reflectivity profiles obtained were fitted with suitable models of electron density contrast to obtain thickness, density and roughness of the individual layers with high resolution (sub nm). The related depth dependent electron scattering length density (eSLD) in units of Å$^{-2}$ which is proportional to the number of atomic components per unit volume can be represented as,

$$\rho_{e_i} = r_e \sum_i N_i Z_i \tag{1}$$

where $r_e$ (=2.818 fm) is the classical electron radius, $N_i$ and $Z_i$ are the depth-dependent number density per unit volume and charge number respectively of the $i^{th}$ component. The electron densities of single layers (i.e. Ni and Zr) and binary (Ni + Zr) interface layers averaged in the sample plane could thus be easily calculated. XRR data analyses for the as-deposited and annealed multilayers are compared in the 3 panels of Fig. 1. The derived eSLD depth profiles corresponding to each multilayer are shown as insets to the XRR data. These profiles provide a physical picture of the effect of annealing on the Ni/Zr multilayers and are useful aids to explain the same.

The R.T. data revealed slight intermixing of Ni and Zr at the interfaces. The constituent layers can be clearly distinguished with the eSLD of Ni being higher than that of Zr, as shown marked in Fig. 1. The multilayer annealed at 200 °C showed significantly higher interfacial mixing than the as-deposited multilayer, wherein both the Ni-on-Zr and Zr-on-Ni interfaces were affected as seen from the eSLD profiles. Comparison of the first order Bragg peak intensities in the XRR data of the R. T. and 200 °C annealed multilayers also revealed a 1.6 times reduction in intensity after annealing. The extent of diffusion can be estimated through calculated values of diffusivity $D$ and diffusion length $L$ given by the following relations (2) and (3):

$$ln[I(t)/I(0)] = -8\pi^2 n^2 Dt/d^2 \tag{2}$$

where $I(0)$ and $I(t)$ represent the Bragg peak intensities obtained from XRR of R.T. and 200 °C annealed multilayers respectively, $n$ is the order of the Bragg peak, $t$ is the annealing time and $d$ is the bilayer width.

$$L = (6Dt)^{0.5} \tag{3}$$

Using Eq. (2), the diffusion constant $D$ describing the diffusivity in the Ni/Zr multilayer annealed to 200 °C was calculated as $1.58 \times 10^{-21}$ m$^2$/sec. This value is of similar order to that reported for Ni diffusion in amorphous Ni-Zr alloys [13,16,35]. The corresponding diffusion length $L$ for this reaction obtained from Eq. (3) was 58 Å.

The top panel of Fig. 1 shows XRR analysis of the multilayer annealed at 400 °C. The first order Bragg peak intensity in this case was observed to have unexpectedly risen to twice the peak intensity of the as-deposited multilayer. This unusual result pointed to a demixing effect that had occurred in the interfacial regions. It is quite obvious from the corresponding eSLD pattern that the Ni-on-Zr interface layers sharpened noticeably while the Ni layers showed elevated eSLD values with considerable reduction in thickness. In contrast the Zr-on-Ni interfaces displayed wide layers, with the thicknesses rising and eSLD values lowering within the depth of the sample. The positions of the Ni-on-Zr and Zr-on-Ni interface layers are marked in the eSLD profile of this figure. The growth of the interfacial layers at the expense of narrowing of the Ni and Zr layers also led to lattice expansion.

PNR measurements characterize both the neutron-nucleus and magnetic interaction for a material. PNR data for the Ni/Zr multilayers were acquired by recording specularly reflected neutron intensities normal to the multilayer surface as a function of momentum transfer $q = (4\pi/\lambda)\sin\theta$ in Å$^{-1}$, where neutron wavelength used was $\lambda = 2.5$ Å and $\theta$ denotes the grazing angles of incidence on the sample surface. As in the case of XRR, here too the relevant structural information of the multilayer could be obtained as a function of depth along the surface normal to the sample. The PNR data were fitted using similar parameters and structural models used to fit the XRR data. These analyses give information on the neutron refractive index depth profile from coherent nuclear scattering length density (nSLD) which can be used to determine chemical composition. The advantage of using polarized neutrons is that the refractive index in this case depends on the relative orientation of the sample in-plane magnetization and the neutron spin. The PNR data can thus yield depth dependent magnetic profile information of a layered sample with high resolution as well. The corresponding nSLD incorporating both chemical and magnetic components is calculated as:

$$\rho_{n_i} = \sum_i N_i \, [ \, b_i \, \pm \, C \, \mu_i ] \tag{4}$$

where $b_i$ is the coherent neutron scattering length and $\mu_i$ is the magnetic moment of the $i^{th}$ component. C = 2.69 fm/$\mu_B$ is a constant. The $\pm$ sign depends on the orientation of sample magnetization with respect to polarization direction of the neutron beam. The neutron reflectivities measured with spin of the neutron beam set parallel (spin-up) and antiparallel (spin-down) to the sample magnetization, are termed as $R^+$ and $R^-$ respectively. Fig. 2a shows the fitted data for PNR measurements made on the as-deposited and annealed Ni/Zr multilayers. The profiles are shown vertically displaced for clarity. The separation between the $R^+$ and $R^-$ traces gives a measure of the extent of magnetization in the sample. The multilayer annealed at 400 °C clearly shows a gap between spin-up and spin-down reflectivities indicating a large increase in magnetic moment with annealing. This aspect is further highlighted in Fig. 2b where the variation of magnetization along the depth for each of the R.T. and 200 °C, 400 °C annealed multilayers are plotted together for comparison.

It can be noted from this figure that there is an obvious difference between the shapes of the Ni-on-Zr and Zr-on-Ni interfaces. When annealed from R.T. to 200 °C the Ni-on-Zr interfaces displayed very little roughness, visible through a slight change of slope of the profiles, indicating that a small amount of mixing had occurred between Ni and Zr atoms. On further annealing to 400 °C, these interfaces considerably sharpened pointing to a trend of demixing. On the other hand, the Zr-on-Ni interfaces underwent steady changes with annealing. After annealing at 400 °C these interfaces became more prominent and flattened out with increase in thickness along the depth. The sharpened Ni-on-Zr interface is shown marked for example in Fig. 2b with a narrow cyan rectangle on the magnetization depth profile obtained at 400 °C. For further reference, the prominent Zr-on-Ni layer formed is indicated on the same profile with a wide yellow rectangle. The overall magnetization appeared to rise considerably as well. There can also be seen a significant rise in the magnetic moment associated with the unreacted Ni layers along with a drastic reduction in their thickness. Average magnetic moment values that could be derived from the PNR analyses of the multilayers at each temperature were as follows: 0.124 μB/atom (at R. T.), 0.165 μB /atom (at 200 °C) and 0.54 μB /atom (at 400 °C). The value of magnetic moment stated for multilayer annealed at 400 °C took into consideration the contributions from both Ni (0.57 μB/atom) and Zr-on-Ni (0.5μB/atom) layers to the same.

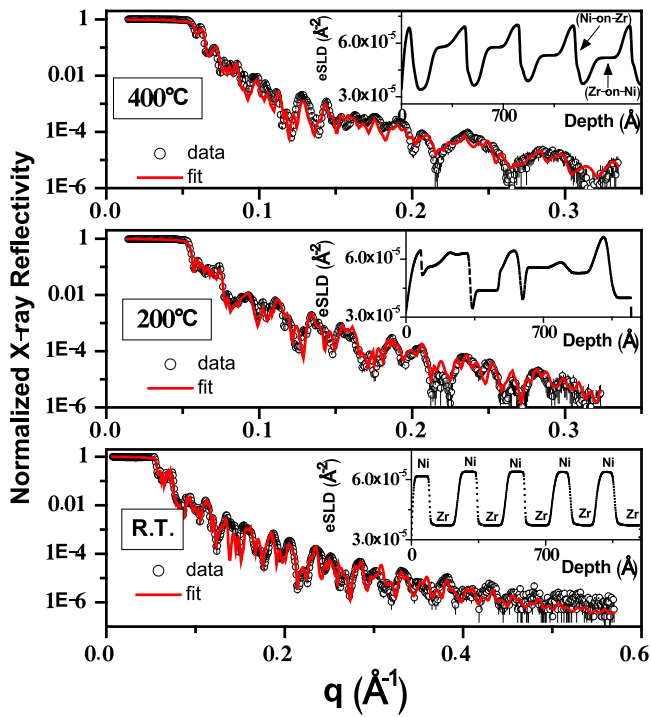

**Fig. 1.** X-ray reflectivity measurements for as-deposited and annealed multilayers. XRR data and profile fits are represented by open circles and solid lines respectively. Insets show the corresponding electron scattering length density variation with depth.

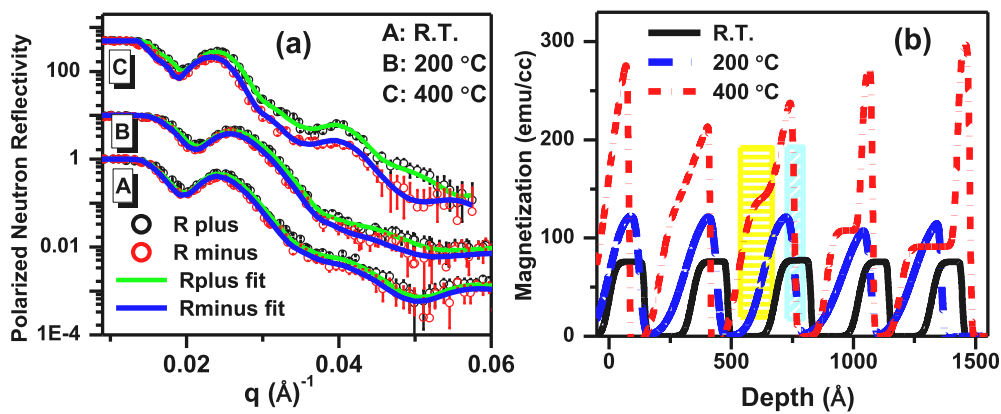

**Fig. 2.** (a) PNR data and corresponding fits for the as-deposited and annealed Ni/Zr multilayers. (b) Derived magnetization depth profiles for each. The positions of the Zr-on-Ni and Ni-on-Zr layers formed at 400 °C are indicated for reference with a wide yellow rectangle and narrow cyan rectangle respectively.

XRR and PNR techniques are non-destructive tools which are used in a complementary manner to probe chemical and magnetic profiles of layered structures and binary materials [36,37]. The average layer thickness and average surface roughness for the Ni, Ni-on-Zr, Zr and Zr-on-Ni layers of the as-deposited and annealed Ni/Zr multilayers obtained from reflectivity data analyses are plotted in Fig. 3. The as-deposited or R.T. data corresponds to 25 °C in these graphs. As annealing proceeded the thickness of the Ni-on-Zr and Zr-on-Ni interface layers rose at the expense of thickness of the Ni and Zr layers, which steadily decreased as expected. It is interesting to note that the trend of reduction in thickness of the original layers with increase in temperature is similar for Ni and Zr layers. This indicates that Zr atoms actively participate in the diffusion across the multilayer interfaces [7] along with Ni atoms, even though their mobility is known to be lesser than that of Ni atoms in this binary system. It was also observed from the reflectivity analyses that the average thickness and density of Ni-on-Zr interface layers in as-deposited and annealed multilayers were less than that of the Zr-on-Ni interface layers. On annealing the as-deposited multilayer to 200 °C, both interfaces showed similar trends of increase in thickness but after 400 °C annealing, the activity of the Zr-on-Ni interface was marked by a large rise in thickness, much higher than the Ni-on-Zr layer. Fig. 3 also shows the changes in average surface roughness of the layers with annealing temperatures. The Ni and Ni-on-Zr layers show a similar trend of initial rise in roughness with annealing up to 200 °C. When annealed at higher temperature, the surfaces of these layers appeared to become smoother. However, the surfaces of Zr-on-Ni interfaces responded to both temperatures with continuous rise in roughness while the Zr layers tended to grow smoother with annealing. Average interfacial roughness of the Ni-on-Zr interfaces is seen to be lesser than that of the Zr-on-Ni interfaces at all temperatures. This aspect can be explained through the atomic size effect during the initial growth modes. Deposition of smaller Ni atoms on amorphous Zr layers during the growth of Ni layers on Zr layers filled the voids between the larger Zr atoms, causing effective smoothening of the interface. This interface behaviour propagated up to the surface of the deposited Ni layer, which can explain the similarity in variation of roughness with annealing for these two layers mentioned earlier. The cause of higher roughness of the Zr-on-Ni interfaces is due to the high energy bombardment of massive Zr atoms during sputtering which disturbed the ordered surface of the crystalline Ni layers. Due to the high mixing entropy of the system, Zr atoms easily penetrated the growing amorphous Zr layer through short circuit paths, in absence of grain boundaries. This is also why the Zr-on-Ni interfaces appeared wider than the Ni-on-Zr interfaces.

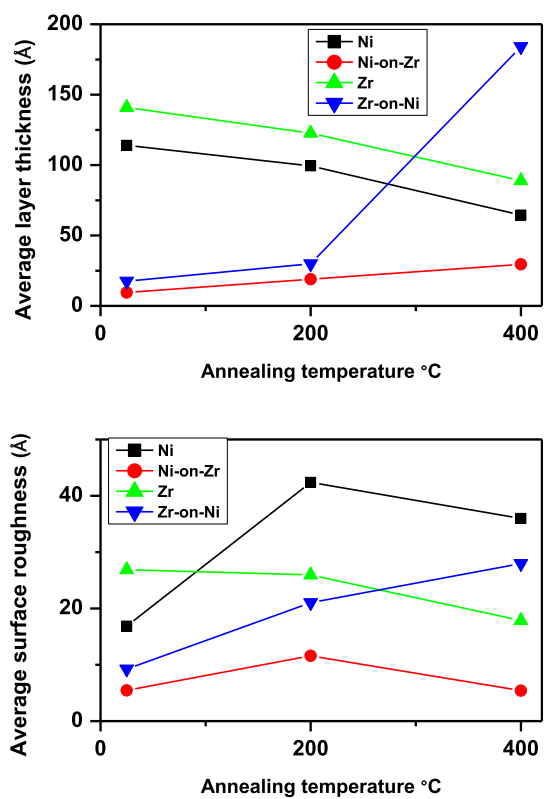

**Fig. 3.** Variation of average values of thickness and surface roughness of the layers with annealing; obtained from reflectivity analyses.

Composition of the mixed interfacial layers in the R.T. and annealed Ni/Zr multilayers were derived through joint analyses of XRR and PNR data using Eqs. (1) and (4) and following the methodology outlined in ref. [36]. The number of atoms/cc of Ni and Zr each in the mixed interface layers of the as-deposited and annealed multilayers derived from these analyses, are graphically represented in the stacked column plots of Fig. 4. The atomic concentrations obtained at each temperature (R.T., 200 °C and 400 °C) are shown coded in the scaled colour map according to the colour key given towards the top of the figure. The abscissa shows 4 layers of the repetitive sequence in the multilayer: unreacted Ni, Ni-on-Zr, unreacted Zr and Zr-on-Ni. The ordinate is scaled in units of $1 \times 10^{23}$ atoms/cc, which also corresponds to the reference values of atomic concentrations of both unreacted Ni and Zr at each temperature. The calculated values of number of atoms/cc of Ni and Zr in the mixed interfacial layers are shown in the segmented columns corresponding to Ni-on-Zr and Zr-on-Ni layers, with temperature rising vertically along the column. For example, the number of atoms/cc of Ni at R.T. is represented by a red block and the green block symbolizes the R.T. number of atoms/cc of Zr. Thus the unreacted Ni layer of the multilayer has a red block corresponding to the reference unit of $1 \times 10^{23}$ atoms/cc and the unreacted Zr layer of the multilayer has a green block corresponding to the reference unit of $1 \times 10^{23}$ atoms/cc. Comparing these values for the interface layers it is seen that the Ni concentration in the Ni-on-Zr layer at R.T. was higher than that in the Zr-on-Ni layer, while the Zr concentration at R.T. showed a reverse trend. This aspect can be explained in the following manner. During the growth of Ni on Zr layer, due to the highly negative enthalpy of mixing in the Ni-Zr system, small Ni atoms rapidly moved into the sparse matrix composed of large Zr atoms and were accommodated in large numbers. In contrast when amorphous Zr was deposited onto crystalline Ni layer, the number of Ni atoms which participated in the interface reaction was commensurate with the number dislodged from the ordered layer by the large sputtered Zr atoms. Moreover, the movement of the mobile Ni atoms was restricted through the growing Zr layer as the latter increased in size. Hence the Ni atoms in this layer were less than those available at the Ni-on-Zr interface.

The Ni atomic ratios were computed as 5.8 for Ni-on-Zr layer and 1.88 for Zr-on-Ni layer. At the ends of the columns depicting the interfacial regions in Fig. 4, are two graphs showing the variation of Ni/Zr atomic ratio with annealing temperature for the respective interface layers. In the Ni-on-Zr layer this ratio is seen to initially increase with annealing and then sharply fall with further annealing. The changes of the Ni/Zr ratio with annealing in the Zr-on-Ni layer were not so marked and in the opposite direction. It is also obvious that the wide differences in Ni and Zr atomic concentrations in the Ni-on-Zr layer as seen from the stacked columns in Fig. 4 do not occur in the Zr-on-Ni layer, wherein the numbers are more similar to each other. The overall value of the Ni/Zr atomic ratio is thus much less in the Zr-on-Ni layer as compared to that of the Ni-on-Zr layer.

Depth distributions of the atomic and molecular species present in the Ni/Zr multilayer annealed at 400 °C were acquired through SIMS measurements. Fig. 5 shows the relevant SIMS profile with traces corresponding to the atomic Ni, Zr and molecular Ni-Zr species varying along the depth of the annealed multilayer. The increased broadening of the peaks observed at higher depths can be attributed to surface roughening during SIMS profiling. The Ni peaks appear sharp and with high intensity. Zr peaks can be seen to reappear at the Ni edges, which points to enhanced intermixing in the Zr-on-Ni region. This is further confirmed by the trace of the Ni-Zr molecular species almost coinciding with the presence of Zr at the interfaces. The peak intensities corresponding to the Ni-Zr molecular species appear to be higher in the Zr-on-Ni interface layer than in the Ni-on-Zr layer as well.

The raw SIMS data was converted into a concentration depth profile using the equation: $C_i = (RSF_i \times I_i) + k_i$, where $C_i$, $RSF_i$ and $I_i$ represent the concentration, relative sensitivity factor and intensity of the $i^{th}$ element respectively. The constant $k_i$ is a weighting factor, which depends on elemental sensitivity and transmission efficiency. The concentrations of Ni and Zr that were determined by this relation, were also normalized by considering $C_{Ni} + C_{Zr} = 100$, where $C_{Ni}$ and $C_{Zr}$ represent the concentrations of Ni and Zr respectively. The average concentration ratios of Ni to Zr estimated from quantitative analyses of SIMS traces revealed a higher value at the Zr-on-Ni interfaces ($C_{Ni}/C_{Zr} = 1.95$) than that calculated for the Ni-on-Zr interfaces ($C_{Ni}/C_{Zr} = 0.35$). The values of these concentration ratios at respective interfaces also matched those obtained through reflectivity analyses.

GIXRD of the as-deposited and annealed multilayers revealed the presence of crystalline Ni only since the Zr layers were amorphous in the R.T. multilayer and continued to remain in that state after annealing. The interfacial mixed layers were observed to be amorphous as well. This is expected from the role of the thermodynamic driving force favouring the lower free energy state in annealed Ni/Zr multilayers [5]. Further, it has also been reported that the nucleation of crystalline phases can take place in an interfacial layer of a Ni/Zr multilayer when the layer thickness is greater than 1000 Å [4]. This is clearly much higher than the range of interfacial layer thicknesses obtained in this study. Fig. 6 compares the development of crystalline Ni peaks in the multilayer with annealing. The intensity of the Ni (1 1 1) peaks increased with annealing while also becoming sharper. The contribution to these peaks arose from both the unreacted Ni layer as well as from the incorporation of Ni crystallites into the Zr-on-Ni interfacial layers during annealing, as seen through analyses of reflectivity data. The prominent Ni (1 1 1) peak was analyzed using Scherrer formula for Ni grain size and effective strain in all the multilayer samples. The grain sizes calculated were as follows, R.T.: 66 Å, 200 °C: 76 Å, 400 °C: 115 Å. The inset shows the variation of both the grain size and strain with annealing. It is seen that as the grain size increased with annealing, the strain in the system reduced.

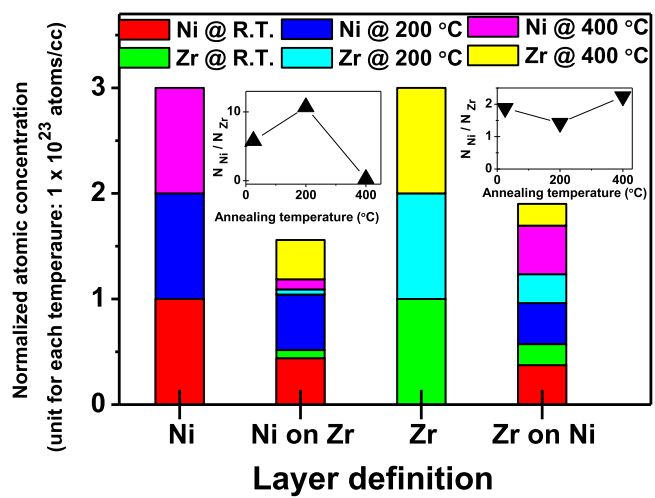

**Fig. 4.** Stacked plots of calculated number of atoms corresponding to Ni and Zr in the interfacial regions Ni-on-Zr and Zr-on-Ni for as-deposited and annealed multilayers. Number of atoms obtained at different temperatures are colour coded.

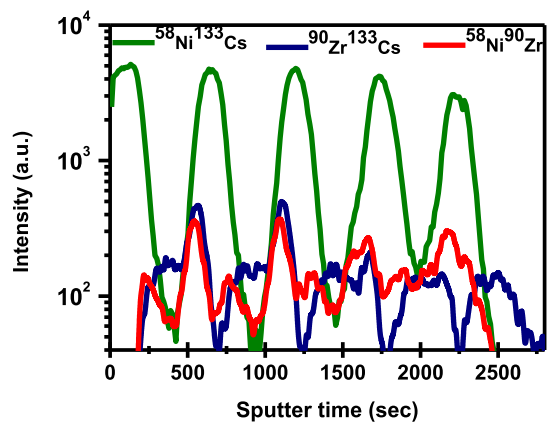

**Fig. 5.** SIMS profile of 400 °C annealed Ni/Zr multilayer in MCs⁺ mode. Variation of Ni and Zr atomic species as well as Ni-Zr molecular species are shown.

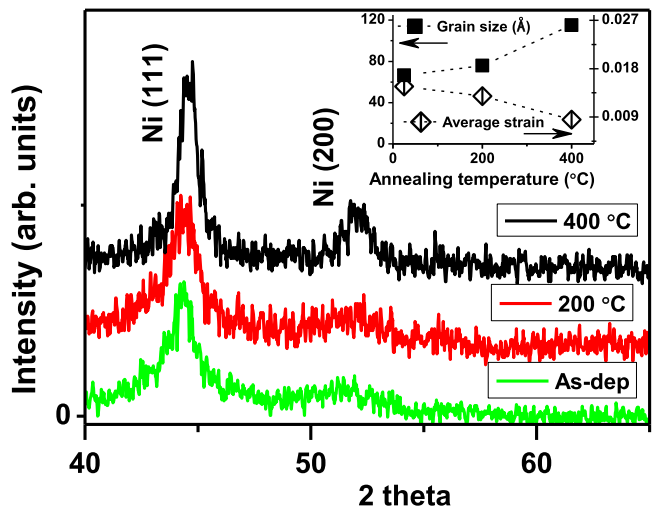

**Fig. 6.** GIXRD plots of as-deposited and annealed multilayers comparing intensity and width of the Ni (1 1 1) peak obtained. Inset shows rise of grain size with annealing and accompanying reduction of strain in the system; calculated from the Ni (1 1 1) peak.

The variation in magnetic properties of the Ni/Zr multilayers as a consequence of annealing was investigated through D.C. magnetization measurements. In Figs. 7 and 8, the changes in magnetization (*M*) with applied field (*H*) recorded for the R.T. and 400 °C annealed multilayers respectively, are displayed. Measurements were made at both 5 K and 300 K. The *M–H* plots for both multilayers exhibited hysteresis, but the shapes of the curves were different. The magnetization plots were fitted with a modified Langevin function to extract the values of saturation magnetization (*M_s*) and average magnetic moment [37]. Other magnetic properties like coercivity (*H_c*) and remanent magnetization (*M_r*) could be obtained directly from the *M–H* curves. The magnetic properties of the multilayers derived from these analyses are listed in Table 1.

The corresponding values of magnetic moments obtained from layer by layer analyses of PNR data for the multilayers have been converted to emu/cc and included in Table 1 for comparison with the $M_s$ values. The high $H_c$ of the R.T. multilayer originates from pure Ni layers with small grains in the as-deposited state. Particles with multiple magnetic domains tend to remain locked, producing large $H_c$, in relation to those with fewer and larger domains. The 400 °C annealed multilayer showed much lower $H_c$. Annealing facilitated the growth of large domains at the expense of smaller ones, which led to easier alignment with the magnetic field giving rise to lower $H_c$ and higher $M_s$. GIXRD analyses of the multilayers also showed increased grain size with annealing. The saturation magnetization obtained after annealing the c-Ni/a-Zr multilayer is about three times that of the as-deposited multilayer. The magnetic moment contributions from the magnetic layers derived through PNR analyses, also match very well with the $M_s$ values obtained from SQUID measurements on the multilayers. The remnant magnetization increased with annealing giving the hysteresis loop a better definition; but the squareness ratio $M_r/M_s$ of both multilayers was close to 0.5, indicating the presence of small single domain particles [38].

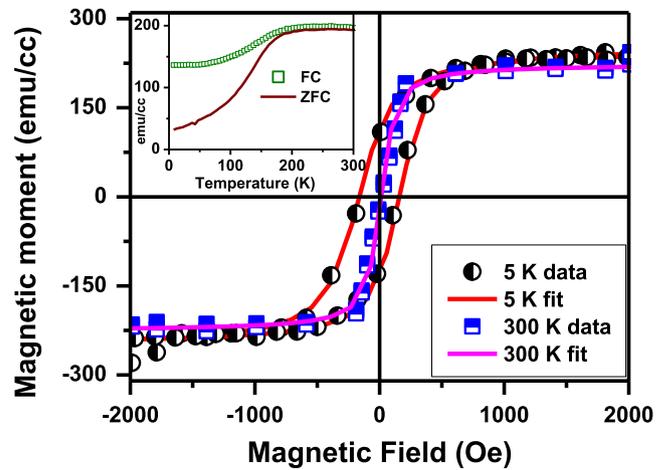

**Fig. 8.** SQUID measurements of 400 °C annealed Ni/Zr multilayer. *M* versus *H* data at 5 K and 300 K have been fitted using Langevin theory. Inset is the variation of *M* with *T* which shows that the blocking temperature from branching of ZFC and FC is very close to R.T.

**Table 1**
Magnetic properties at 300 K obtained from SQUID and PNR data analyses.

|  | $H_c$ (Oe) | $M_r$ (emu/cc) | $M_s$ (emu/cc) | Magnetic moment from PNR analyses (emu/cc) |
|---|---|---|---|---|
| R.T. | 300 | 31 | 73 | 77 |
| 400 °C | 161 | 115 | 225 | 205 |

The insets in Figs. 7 and 8 are the magnetization versus temperature (*M–T*) curves for the R.T. and 400 °C multilayers respectively. Measurements were made in both field cooled (FC) and zero field cooled (ZFC) modes. The shapes of these curves are effectively governed by a change in magnetic anisotropy of the system. A measure of this value is given by the magnetic anisotropy constant defined as $K = 25 \ k_B \ T_b \ V^{-1}$ [39], where $k_B$ is the Boltzmann constant, $T_b$ is the blocking temperature occurring at the branching of FC and ZFC curves, $V$ is the volume of a single particle. Magnetic anisotropy values were calculated for the multilayers as: $K_{R.T.} = 2.3\text{E}06 \ \text{erg/cm}^3$ and $K_{400 \ °C} = 5.7\text{E}05 \ \text{erg/cm}^3$. These values are higher than $K$ of bulk Ni, and this can be attributed to the nanoparticle sizes obtained for the multilayer [40], with $K$ decreasing for increasing grain size [41]. Annealing had thus reduced the anisotropy in the system. $H_c$ values are also determined through the energetic contributions to magnetic anisotropy. In the R.T. multilayer, $K$ is high and so in this case spins are initially aligned in preferred directions. The applied field is then insufficient to align the spins in the field direction due to which the resultant magnetization is small and $H_c$ is high. In order to generate ZFC, no field is applied during cooling hence spins can be frozen in random directions and the ZFC curve decays as shown in the inset of Fig. 7. After annealing, the multilayer exhibited low K implying that the randomness had reduced. Hence spins could be easily aligned to the applied field direction when the sample was cooled during FC magnetization (low $H_c$). The inset of Fig. 8 shows that the FC curve thus remains almost constant with decreasing temperature and ZFC decreases slightly with a broad maximum.

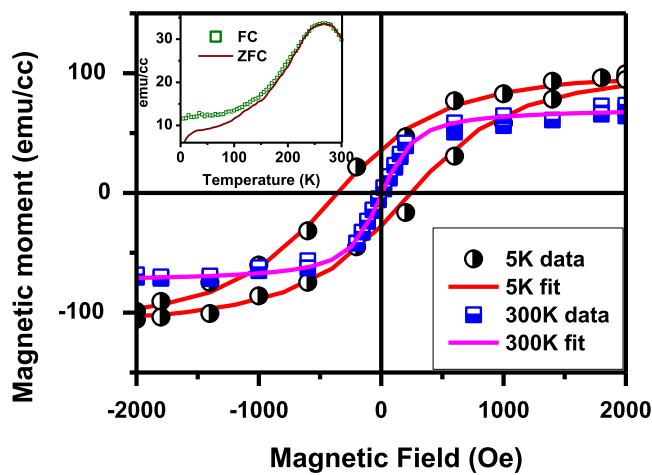

**Fig. 7.** SQUID measurements of as-deposited Ni/Zr multilayer. *M* versus *H* data recorded at 5 K and 300 K have been fitted using Langevin theory. Inset shows plot of *M* versus *T* and the branching of ZFC and FC originating around R.T.

## 4. Discussion

The results of this study imply that the movements of Ni and Zr atoms across the interfaces of the Ni/Zr multilayers as a function of annealing are governed by both kinetic constraints and thermodynamical aspects. It was also observed that the Ni-on-Zr and Zr-on-Ni interfaces behaved very differently with annealing. In Fig. 9 the behaviour of a repetitive section of the Ni/Zr multilayer is represented through a scheme, summarizing the atomic diffusion occurring within the Ni/Zr multilayers with annealing as observed in this study.

Annealing of the R.T. multilayer to 200 °C caused Ni atoms to move in large numbers into both Ni-on-Zr and Zr-on-Ni interface layers; which were already formed at R.T. with the driving force being provided by the mixing entropy of the Ni-Zr system. This led to a reduction in thickness of the Ni layers and an increase in thickness of both interface layers. The calculated value of $D$ describing the diffusion for this route was on par with that reported in literature for annealed Ni/Zr multilayers. The value of $L$ calculated from Eq. (3) was approximately equal to the sum of the average interfacial layer thicknesses obtained after 200 °C annealing; taken from Fig. 3. However due to the rapid diffusion of Ni into the low-density Ni-on-Zr layer at 200 °C, approximately 35% of Zr atoms from this interface formed at R.T. flowed back into the Zr layer. The Ni content at the Ni-on-Zr interface was thus more than that in Zr-on-Ni layer and the reverse was observed for the Zr atoms at this temperature. These variations were similar to that noted for the R.T. multilayer, as explained earlier in this work. The corresponding Ni/Zr atomic ratios at 200 °C were calculated as: 10.7 for the Ni-on-Zr layer and 1.4 for the Zr-on-Ni layer (Fig. 4). When the multilayer was further annealed to 400 °C, Ni atoms were observed to largely deplete from the Ni-on-Zr interface to the extent of 82% back to the Ni layer. The Ni/Zr atomic ratio of this layer was then found to have drastically dropped to 0.26. This was also corroborated by SIMS measurements. Since Zr atoms from the Zr layer continued to move into this interface, the thickness of this interface layer increased slightly beyond that obtained after annealing at 200 °C.

On the other hand, the movement of Ni atoms into the Zr-on-Ni interface at 400 °C led to a large rise in thickness and a slight drop in the number of Zr atoms/cc by 25%, as they demixed from the interface layer and moved into the amorphous Zr layer. As a result, the ratio of Ni to Zr atoms rose to 2.24 in this layer. This was also confirmed through quantitative analyses of the SIMS profile of 400 °C annealed multilayer. Meanwhile the density of the Ni layer increased considerably despite the high flow of Ni atoms into the Zr-on-Ni interface layer; enormously raising the value of magnetic moment from both these layers. It was deduced from these observations that the composition and thickness of interface layers in a crystalline/amorphous multilayer subjected to thermal treatments can vary in a certain way. This is coupled with the important fact of stability of the mixed layers. The supersaturated Ni/Zr atomic ratio of 10.7 obtained in the Ni-on-Zr layer could obviously not be sustained within the narrow confines of the restricted volume in the multilayer, leading to demixing of the more mobile species.

Several simulation studies in binary fluid mixtures using the hard-sphere approximation have established certain conditions under which demixing in such systems due to thermal treatments can occur [42,43]. Although these studies do not strictly pertain to solid film systems, it is interesting to note that some of the key features defined for likely demixing in binary systems draw parallels with those of the Ni-Zr system. One of the required features stated that the constituent atoms must possess a wide atomic size difference, which is also true for the early-late transition metal couple Ni-Zr, engendering its limited solid solubility. However, the discussion in the mentioned studies was related to the ensuing excess volume in the system; the prompt mixing of Ni and Zr in the Ni-Zr system being driven by a highly negative enthalpy of mixing. Another aspect of the likely candidates for demixing in a binary system which was brought out by the same simulation studies was that, the attraction between dissimilar atoms exceeds that between like atoms. This is pertinent to the Ni-Zr system as well [36]. In the present study, the demixing of atoms in the Ni-on-Zr layer caused the enthalpy of the system to drop. In order to achieve a stable state, this was balanced by a rise in configurational entropy contributed by the large number of Ni atoms getting redistributed in the Ni layer. Segregation of nanocrystalline Ni towards the top of an annealed Ni-Zr alloy thin film for the sake of reducing free energy of the overall system has been reported earlier [37]. It is interesting to note that the electrocatalytic activity of amorphous Ni-Zr alloys for hydrogen evolution applications can be enhanced through the presence of such a Ni layer on top [44]. The close packing of the Ni atomic planes reduced the interfacial free energy in the 400 °C annealed multilayer which contributed to more stability in the system; as seen by lattice strain in the system being relieved as the Ni grain sizes increased with annealing (Fig. 6). The Ni-on-Zr layers thus underwent demixing at both temperatures causing reduction in both their thickness

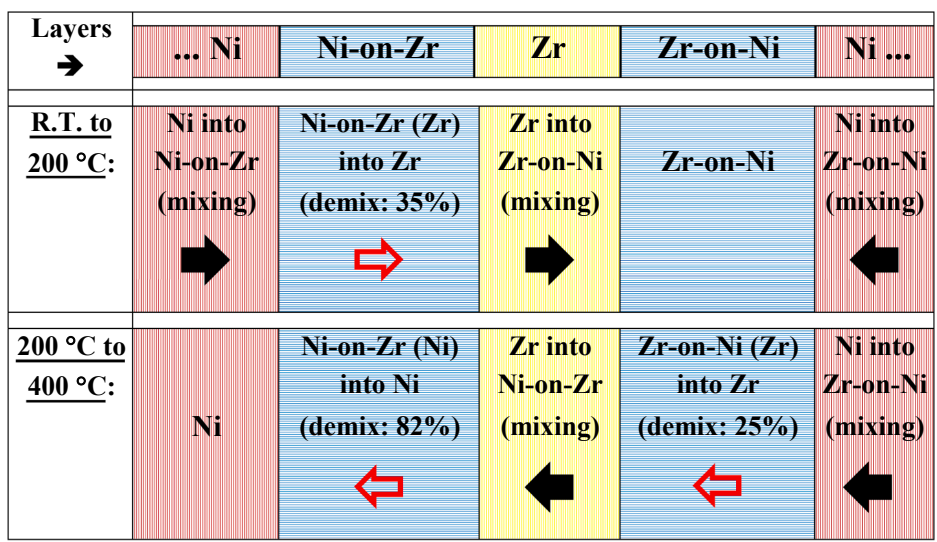

| Layers → | ... Ni | Ni-on-Zr | Zr | Zr-on-Ni | Ni ... |
|---|---|---|---|---|---|
| **R.T. to 200 °C:** | Ni into Ni-on-Zr (mixing) → | Ni-on-Zr (Zr) into Zr (demix: 35%) ⇒ | Zr into Zr-on-Ni (mixing) → | Zr-on-Ni | Ni into Zr-on-Ni (mixing) ← |
| **200 °C to 400 °C:** | Ni | Ni-on-Zr (Ni) into Ni (demix: 82%) ⇐ | Zr into Ni-on-Zr (mixing) ← | Zr-on-Ni (Zr) into Zr (demix: 25%) ⇐ | Ni into Zr-on-Ni (mixing) ← |

**Fig. 9.** A scheme depicting the flow of Ni and Zr atoms across interfaces within a repetitive section of the Ni/Zr multilayer, as a consequence of thermal annealing. Black solid arrows indicate mixing of components in the direction of the arrow. Red outlined arrows specify demixing of atoms from that layer.

and density, as compared to those of the Zr-on-Ni layers.

The magnetic moment per unit volume of the as-deposited and annealed c-Ni/a-Zr multilayers obtained through measurements from two different techniques, D.C. magnetization and PNR, corroborated each other very well (Table 1). The elevation in magnetic moment after annealing at 400 °C was attributed to both the considerable densification in the narrow Ni layer and incorporation of Ni crystallites into the Zr-on-Ni layer. This was accompanied by a decrease in magnetic anisotropy. The effect of annealing in this system was therefore to produce a soft magnetic material with low anisotropy, characterized by large saturation magnetization and small coercivity. The latter is also associated with large grain sizes, which were obtained after annealing the c-Ni/a-Zr multilayer and seen by GIXRD analyses.

## 5. Summary and conclusions

Magnetron sputter deposited multilayers composed of 5 bilayers of crystalline Ni/amorphous Zr were vacuum annealed at 200 °C and 400 °C to observe the compositional variations at the Ni-on-Zr and Zr-on-Ni interfaces. Both the interfacial layers responded actively to annealing through atomic diffusion across the interfaces, though the effect was more pronounced at higher temperature. The Ni-on-Zr interface layer was found to be unstable at both temperatures; with Zr atoms diffusing out at 200 °C and a large number of Ni atoms diffusing out at 400 °C. These atomic movements caused the Ni/Zr atomic ratio to drastically fall at this interface. As a consequence of annealing, the Zr-on-Ni layer was steadily populated with Ni atoms at both temperatures while Zr diffused out to a small extent at 400 °C. The observed demixing phenomena contributed to reducing the overall free energy of the system in order to ultimately attain a thermodynamically stable configuration. A large increase in magnetic moment obtained after annealing the c-Ni/a-Zr multilayers at 400 °C, was attributed to the high densification of the Ni layers as well as to the incorporation of Ni crystallites into the Zr-on-Ni interface layers. The effect of annealing these multilayers was to lower the value of magnetic anisotropy and produce a soft magnetic material.

It was concluded from this study that the deposition of a crystalline layer on an amorphous layer in a binary multilayer composed of miscible elements, produced a thermally unstable interface. In contrast, the interface formed in between an amorphous layer grown on a crystalline layer responded to high temperature annealing through accelerated growth of the interface width. The results of this study on thermally induced interfacial behaviour in an amorphous/crystalline Ni/Zr multilayer will be useful in determining the properties and performance of future composite materials. These findings on the diffusional aspects of Ni-Zr system can be used to modify these materials for better hydrogen separation and high strength structural applications. Control over the thermally induced interfacial behaviour in magnetic/non-magnetic multilayers can also tailor the formation of novel materials with technologically appropriate magnetic properties.

## CRediT authorship contribution statement

**Debarati Bhattacharya:** Conceptualization, Methodology, Validation, Formal analysis, Investigation, Writing – original draft, Writing – review & editing, Visualization. **Vijay Karki:** Methodology, Formal analysis. **Surendra Singh:** Software, Resources, Data curation. **T.V. Chandrasekhar Rao:** Investigation, Supervision.

## Declaration of Competing Interest

The authors declare that they have no known competing financial interests or personal relationships that could have appeared to influence the work reported in this paper.

## Acknowledgements:

The author thanks Swapan Jana and Pooja Moundekar for their technical assistance during multilayer deposition and X-ray measurements. The author is grateful for the insightful comments of the reviewer which helped enrich the manuscript favourably.